\documentclass[conference, compsocconf]{IEEEtran}
\usepackage{amsmath}
\usepackage{amssymb}
\usepackage{color}
\usepackage{supertabular,booktabs}
\usepackage{algorithm}
\usepackage{algorithmic}
\usepackage{graphicx}
\usepackage{bm}
\usepackage{booktabs}%三线表
\usepackage{multirow}
\usepackage{mathrsfs}
\usepackage[colorlinks,linkcolor=red,anchorcolor=blue,citecolor=green,CJKbookmarks=True]{hyperref}

\begin{document}
\title{Adversarial Deep Reinforcement Learning in Portfolio Management}
\author
{Zhipeng Liang \IEEEauthorrefmark{1}\IEEEauthorrefmark{2},\IEEEauthorblockN{Hao Chen\IEEEauthorrefmark{1}\IEEEauthorrefmark{2}, Junhao Zhu\IEEEauthorrefmark{1}\IEEEauthorrefmark{2}, Kangkang Jiang \IEEEauthorrefmark{1}\IEEEauthorrefmark{2},Yanran Li \IEEEauthorrefmark{1}\IEEEauthorrefmark{2}
}
\IEEEauthorblockA
{
	\IEEEauthorrefmark{1}Likelihood Technology\\
}
\IEEEauthorblockA
{
	\IEEEauthorrefmark{2}Sun Yat-sen University\\
}
$ $\\
$\{liangzhp6,chenhao348,zhujh25,jiangkk3,liyr8\}@mail2.sysu.edu.cn$
}

\maketitle
\begin{abstract}
 In this paper, we implement three state-of-art continuous reinforcement learning algorithms, Deep Deterministic Policy Gradient (DDPG), Proximal Policy Optimization (PPO) and Policy Gradient (PG)in portfolio management. All of them are widely-used in game playing and robot control. What's more, PPO has appealing theoretical propeties which is hopefully potential in portfolio management. We present the performances of them under different settings, including different learning rates, objective functions, feature combinations, in order to provide insights for parameters tuning, features selection and data preparation. We also conduct intensive experiments in China Stock market and show that PG is more desirable in financial market than DDPG and PPO, although both of them are more advanced. What's more, we propose a so called Adversarial Training method and show that it can greatly improve the training efficiency and significantly promote average daily return and sharpe ratio in back test. Based on this new modification, our experiments results show that our agent based on Policy Gradient can outperform UCRP.

\end{abstract}


\IEEEpeerreviewmaketitle

\section{Introduction}
Utilizing deep reinforcement learning in portfolio management is gaining popularity in the area of algorithmic trading. However, deep learning is notorious for its sensitivity to neural network structure, feature engineering and so on. Therefore, in our experiments, we explored influences of different optimizers and network structures on trading agents utilizing three kinds of deep reinforcement learning algorithms, deep deterministic policy gradient (DDPG), proximal policy optimization (PPO) and policy gradient (PG). Our experiments were conveyed on datasets of China stock market. Our codes can be viewed on github\footnote{https://github.com/qq303067814/Reinforcement-learning-in-portfolio-management-}.

\section{Summary}
This paper is mainly composed of three parts. First, portfolio management, concerns about optimal assets allocation in different time for high return as well as low risk. Several major categories of portfolio management approaches including "Follow-the-Winner", "Follow-the-Loser", "Pattern-Matching" and "Meta-Learning Algorithms" have been proposed. Deep reinforcement learning is in fact the combination of "Pattern-Matching" and "Meta-Learning" [\ref{Online portfolio selection}].

Reinforcement learning is a way to learn by interacting with environment and gradually improve its performance by trial-and-error, which has been proposed as a candidate for portfolio management strategy. Xin Du et al. conducted Q-Learning and policy gradient in reinforcement learning and found direct reinforcement algorithm (policy search) enables a simpler problem representation than that in value function based search algorithm [\ref{Q&RRL}]. Saud Almahdi et al. extended recurrent reinforcement learning and built an optimal variable weight portfolio allocation under the expected maximum drawdown [\ref{EMDD}]. Xiu Gao et al. used absolute profit and relative risk-adjusted profit as performance function to train the system respectively and employ a committee of two network, which was found to generate appreciable profits from trading in the foreign exchange markets [\ref{Sharpe Ratio Maximization}].

Thanks to the development of deep learning, well known for its ability to detect complex features in speech recognition, image identification, the combination of reinforcement learning and deep learning, so called deep reinforcement learning, has achieved great performance in robot control, game playing with few efforts in feature engineering and can be implemented end to end [\ref{DRL}]. Function approximation has long been an approach in solving large-scale dynamic programming problem [\ref{PG with appro}].  Deep Q Learning, using neural network as an approximator of Q value function and replay buffer for learning, gains remarkable performance in playing different games without changing network structure and hyper parameters [\ref{DQN}]. Deep Deterministic Policy Gradient(DDPG), one of the algorithms we choose for experiments, uses actor-critic framework to stabilize the training process and achieve higher sampling efficiency [\ref{DDPG}]. Another algorithm, Proximal Policy Optimization(PPO), turns to derive monotone improvement of the  policy [\ref{PPO}].

Due to the complicated, nonlinear patterns and low signal noise ratio in financial market data, deep reinforcement learning is believed potential in it. Zhengyao Jiang et al. proposed a framework for deep reinforcement learning in portfolio management and  demonstrated that it can outperform conventional portfolio strategies [\ref{DRLPM}]. Yifeng Guo el at. refined log-optimal strategy and combined it with reinforcement learning [\ref{RLORL}]. Lili Tang proposed a model-based actor-critic algorithm under uncertain
environment is proposed, where the optimal value function is obtained by iteration on the basis of the constrained risk
range and a limited number of funds [\ref{AC}]. David W. Lu implemented in Long Short Term Memory (LSTM) recurrent
structures with Reinforcement Learning or Evolution Strategies acting as agents The robustness and feasibility of the system is
verified on GBPUSD trading [\ref{LSTM}]. Steve Y. Yang et al. proposed an investor sentiment reward based trading system aimed at extracting only signals that generate either negative or positive market responses [\ref{inverse}]. Hans Buehler presented a framework for hedging a portfolio of derivatives in the presence of market frictions such as transaction costs, market impact, liquidity constraints or risk limits using modern deep reinforcement machine learning methods [\ref{DeepHedging}].

However, most of previous works use stock data in America, which cannot provide us with implementation in more volatile China stock market. What's more, few works investigated the influence of the scale of portfolio or combinations of different features. To have a closer look into the true performance and uncover pitfalls of reinforcement learning in portfolio management, we choose mainstream algorithms, DDPG, PPO and PG and do intensive experiments using different hyper parameters, optimizers and so on.

The paper is organized as follows: in the second section we will formally model portfolio management problem. We will show the existence of transaction cost will make the problem from a pure prediction problem whose global optimized policy can be obtained by greedy algorithm into a computing-expensive dynamic programming problem. Most reinforcement learning algorithms focus on game playing and robot control, while we will show that some key characters in portfolio management requires some modifications of the algorithms and propose our novel modification so called Adversarial Training. The third part we will go to our experimental setup, in which we will introduce our data processing, our algorithms and our investigation into effects of different hyper parameters to the accumulated portfolio value. The fourth part we will demonstrate our experiment results. In the fifth part we would come to our conclusion and future work in deep reinforcement learning in portfolio management.

\section{Problem Definition}
 Given a period, e.g. one year, a stock trader invests into a set of assets and is allowed to reallocate in order to maximize his profit. In our experiments, we assume that the market is continuous, in other words, closing price equals open price the next day. Each day the trading agent observes the stock market by analyzing data and then reallocates his portfolio. In addition, we assume that the agent conducts reallocation at the end of trade days, which indicates that all the reallocations can be finished at the closing price. In addition, transaction cost, which is measured as a fraction of transaction amount, has been taken into considerations in our experiments.

 Formally, the portfolio consists of m+1 assets, including m risky assets and one risk-free asset. Without depreciation, we choose money as the risk-free asset. The closing price of $i^{th}$ asset after period t is $v^{close}_{i,t}$. The closing price of all assets comprise the price vector for period t as $v^{close}_t$. Modeling as a Markovian decision process, which indicates the next state only depends on current state and action. Tuple $(S,A,P,r,\rho_0,\gamma)$ describes the entire portfolio management problem where $S$ is a set of states, $A$ is a set of actions, $P:S\times A \times S \rightarrow \mathbb{R}$ is the transition probability distribution, $r:S \rightarrow \mathbb{R}$ is the reward function. $\rho_0:S\rightarrow \mathbb{R}$ is the distribution of the initial state $s_0$ and $\gamma \in (0,1)$ is the discount factor.

 It's worth to note that in Markovian decision process, most objective functions take the form of discount rate, which is $R=\sum_{t=1}^{T}\gamma^tr(s_t,a_t)$. However, in the area of portfolio management, due to the property that the wealth accumulated by time t would be reallocated in time t+1, indicating that the wealth at time T, $P_T=\prod_{t=1}^{T}P_0 r_t$ is continued product form but not summation. A sightly modification would be needed, which is to take logarithm of the return to transform continued product form into summation.

 To clarify each item in the Markovian decision process, we make some notations here. Define $y_{t}=\frac{\mathbf{v_t}}{\mathbf{v_{t-1}}}=(1,\frac{v_{1,t}}{v_{1,t-1}},\dots,\frac{v_{m,t}}{v_{m,t-1}})^T$as the price fluctuating vector. $\mathbf{w_{t-1}}=(w_{0,t-1},w_{1,t-1},\dots,w_{m,t-1})^T$ represents the reallocated weight at the end of time $t-1$ with constraint $\sum_i w_{i,t-1}=1$. We assume initial wealth is $P_0$. Definitions of state, action and reward in portfolio management are as below.

 \begin{itemize}
 	\item State($s$): one state includes previous open, closing, high, low price, volume or some other financial indexes in a fixed window.
 	\item Action($a$): the desired allocating weights, $a_{t-1}=(a_{0,t-1},a_{1,t-1},\dots,a_{m,t-1})^T$ is the allocating vector at period $t-1$, subject to the constraint $\sum_{i=0}^{n}a_{i,t-1}=1$. Due to the price movement in a day, the weights vector $a_{t-1}$ at the beginning of the day would evolve into $w_{t-1}$ at the end of the day:
 	$$\mathbf{w_{t-1}}=\frac{\mathbf{y_{t-1}}\odot \mathbf{a_{t-1}}}{\mathbf{y_{t-1}} \cdot \mathbf{a_{t-1}}}$$
 	
 	\begin{figure}[ht]
 		\centering
 		\includegraphics[scale=0.29]{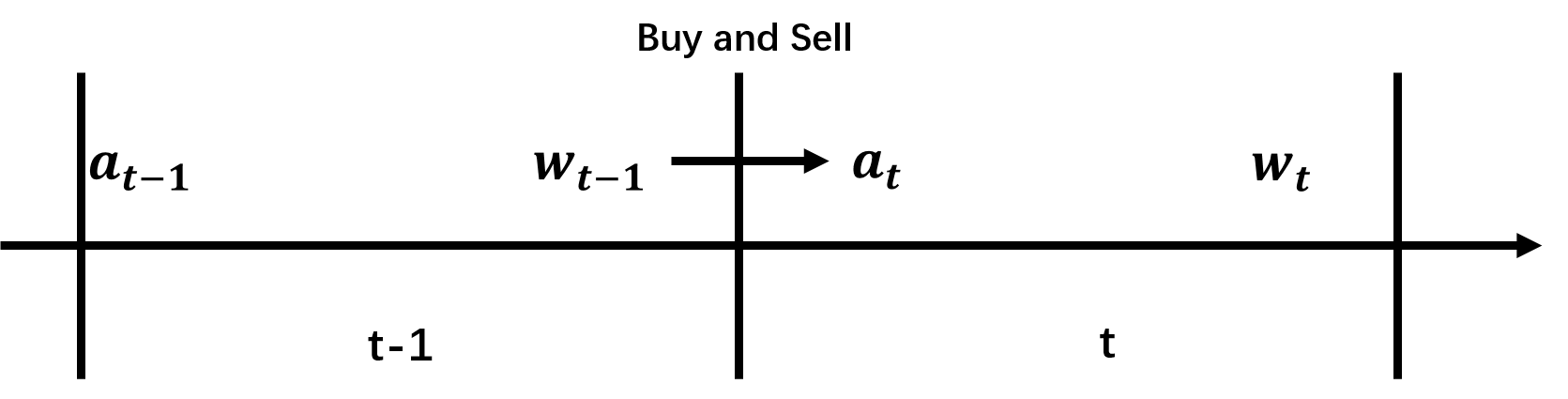}
 		\caption{The evolution of weights vector} \label{fig 5}
 	\end{figure}
 	
 	\item Reward($r$): the naive fluctuation of wealth minus transaction cost. The fluctuation of wealth is  $a_{t-1}^T \cdot y_{t-1}$. In the meanwhile, transaction cost should be subtracted from that, which equals $\mu \sum_{i=1}^{m}|a_{i,t-1}-w_{i,t-1}|$. The equation above suggests that only transactions in stocks occur transaction cost. Specifically, we set $\mu=0.25\%$. In conclusion, the immediate reward at time t-1 as:
 	$$r_t(s_{t-1},a_{t-1})=\log(\mathbf{a_{t-1}} \cdot \mathbf{y_{t-1}}-\mu \sum_{i=1}^{m}|a_{i,t-1}-w_{i,t-1}|)$$.
 	
 \end{itemize}

 The introduction of transaction cost is a nightmare to some traditional trading strategy, such as follow the winner, follow the loser etc. Even can we predict precisely all stock price in the future, deriving the optimal strategy when the period is long or the scale of portfolio is large, is still intractable. Without transaction cost, greedy algorithm can achieve optimal profits. To be specific, allocating all the wealth into the asset which has the highest expected increase rate is the optimal policy in such a naive setting. However, the existence of transaction cost might turn action changing too much from previous weight vector into suboptimal action if the transaction cost overweights the immediate return.

Although rich literatures have discussed Markovian decision process, portfolio management is still challenging due to its properties. First and foremost, abundant noise included in the stock data leads to distorted prices. Observations of stock prices and financial indexes can hardly reflect the states underneath. Providing inefficient state representations for the algorithm would lead to disastrous failure in its performance.

 What's more, the transition probability of different states is still unknown. We must learn environment before we attempt to solve such a complex dynamic programing problem.

 Although buying and selling stocks must be conducted by hands, here we still adapt continuous assumption. In fact when wealth is much more than the prices of stocks, such a simplification would not lose much generation.

 \section{Deep Reinforcement learning}
 Reinforcement learning, especially combining with state-of-art deep learning method is therefore thought to be a good candidate for solving portfolio problem. Reinforcement learning is a learning method, by which the agent interacts with the environment with less prior information and learning from the environment by trail-and-error while refining its strategy at the same time. Its low requirements for modeling and feature engineering is suitable for dealing with complex financial markets. What's more, deep learning has witnessed its rapid progress in speech recognition and image identification. Its outperformance with conventional methods has proven its capability to capture complex, non-linear patterns. In fact, different methods using neural network in designing trading algorithms have been proposed.

 Compared with solely using deep learning or reinforcement learning in portfolio management, deep reinforcement learning mainly has three strengths.

 First, with market's information as its input and allocating vector as its output, deep reinforcement learning is an totally artificial intelligent methods in trading, which avoids the hand-made strategy from prediction of the future stock price and can fully self-improved.

 Second, deep reinforcement learning does not explicitly involve predictions towards stock performance, which has been proven very hard. Therefore, less challenges would hinder the improvement in reinforcement learning performance.

 Third, compared with conventional reinforcement learning, deep reinforcement learning approximates strategy or value function by using neural network, which can not only include the flexibility of designing specific neural network structure but also prevent so called "curse of dimensionality", enabling large-scale portfolio management.

 Several continuous reinforcement learning methods have been proposed, such as policy gradient, dual DQN, Deep Deterministic Policy Gradient and Proximal Policy Optimization. We conduct the latter two algorithms in our experiments to test their potential in portfolio management.

 \subsection{Deep Deterministic Policy Gradient}
 Deep Deterministic Policy Gradient(DDPG) is a combination of Q-learning and policy gradient and succeed in using neural network as its function approximator based on Deterministic Policy Gradient Algorithms [\ref{DPG}]. To illustrate its idea, we would briefly introduce Q-learning and policy gradient and then we would come to DDPG.

 Q-learning is a reinforcement learning based on Q-value function.
 To be specific, a Q-value function gives expected accumulated reward when executing action \emph{a} in state \emph{s} and follow policy $\pi$ in the future, which is:
 $$Q^{\pi}(s_t,a_t)=\mathbb{E}_{r_i\geq t,s_i> t~E,a_i>t~\pi}[R_t|s_t,a_t]
 $$

The Bellman Equation allows us to compute it by recursion:
  $$Q^{\pi}(s_t,a_t)=\mathbb{E}_{r_t,s_{t+1} \sim E}[r(s_t,a_t)+\gamma \mathbb{E}_{a_{t+1}\sim \pi}[Q^{\pi}(s_{t+1},a_{t+1})]]
 $$

 For a deterministic policy which is a function $\mu:S\rightarrow A$, the above equation can be written as:
 $$Q^{\pi}(s_t,a_t)=\mathbb{E}_{r_t,s_{t+1} \sim E}[r(s_t,a_t)+\gamma[Q^{\mu}(s_{t+1},\mu(s_{t+1}))]]
 $$

 To be specific, Q-learning adapts greedy policy which is:
 $$\mu(s)=\arg \max_{a}Q(s,a)$$

Deep reinforcement learning uses neural network as the Q-function approximator and some methods including replay buffer are proposed to improve the convergence to the optimal policy. Instead of using iterations to derive the  conventional Q-value function, the function approximator, parameterized by $\theta^Q$, is derived by minimizing the loss function below:
$$L(\theta^Q)=\mathbb{E}_{s_t\sim \rho^\beta, a_t\sim \beta, r_t\sim }[(Q(s_t,a_t|\theta^Q)-y_t)^2]$$
where
$$y_t=r(s_t,a_t)+\gamma Q(s_{t+1},\mu(s_{t+1})|\theta^Q)$$

It's worth to note here that $y_t$ is calculated by a separate target network which is softly updated by online network. This simple change moves the relatively unstable problem of
learning the action-value function closer to the case of supervised learning, a problem for which
robust solutions exist. This is another method to improve convergence.

When dealing with continuous action space, naively implementing  Q-learning is intractable when the action space is a large due to the "curse of dimensionality". What's more, determining the global optimal policy in an arbitrary Q-value function may be infeasible without some good features guaranteed such as convex.

The answer of DDPG to address the continuous control problem is to adapt policy gradient, in which DDPG consists of an actor which would directly output continuous action. Policy would then be evaluated and improved according to critic, which in fact is a Q-value function approximator to represent objective function. Recall the goal of Markovian decision process: derive the optimal policy which maximize the objective function. Parameterized by $\theta$, we can formally write it as:

\begin{equation*}
	\begin{aligned}
	\tau&=(s_1,a_1,s_2,a_2,\dots)\\
	J(\pi_\theta)&=\mathbb{E}_{\tau \sim p_\theta(\tau)}[\sum_{t}\gamma^t r(s_t,a_t)]\\
		\pi_{\theta^{*}}
		&=\arg \max_{\pi_\theta}J(\pi_\theta)\\
		&=\arg \max_{\pi_\theta}\mathbb{E}_{\tau \sim p_\theta(\tau)}[\sum_{t}\gamma^t r(s_t,a_t)]\\
		&=\arg \max_{\pi_\theta}\mathbb{E}_{\tau \sim p_\theta(\tau)}[r(\tau)]\\
		&=\arg \max_{\pi_\theta}\int \pi_\theta(\tau)r(\tau)d\tau
	\end{aligned}
\end{equation*}

In deep reinforcement learning, gradient descent is the most common method to optimize given objective function, which is usually non-convex and high-dimensional.
Taking derivative of the objective function equals to take derivative of policy. Assume the time horizon is finite, we can write the strategy in product form:
\begin{equation*}
\begin{aligned}
\pi_\theta(\tau)&=\pi_\theta(s_1,a_1,\dots,s_T,a_T)\\
&=p(s_1)\prod_{t=1}^{T}\pi_\theta(a_t|s_t)p(s_{t+1}|s_t,a_t)
\end{aligned}
\end{equation*}

However, such form is difficult to make derivative in terms of $\theta$. To make it more computing-tractable, a transformation has been proposed to turn it into summation form:
\begin{equation*}
	\begin{aligned}
	\nabla_{\theta}\pi_\theta(\tau)&=\pi_\theta(\tau)\frac{\nabla_\theta \pi_\theta(\tau)}{\pi_\theta(\tau)}\\
	&=\pi_\theta(\tau)\nabla_\theta \log \pi_\theta(\tau)\\	
	\nabla_\theta \log \pi_\theta(\tau)&=\nabla_{\theta}(\log p(s_1)+\sum_{t=1}^{T}\log \pi_\theta(a_t|s_t)+\log p(s_{t+1}))\\
	&=\sum_{t=1}^{T}\nabla_\theta \log \pi_\theta(a_t,s_t)
	\end{aligned}
\end{equation*}

Therefore, we can rewrite differentiation of the objective function into that of logarithm of policy:
\begin{equation*}
	\begin{aligned}
		\nabla J(\pi_\theta)&=\mathbb{E}_{\tau \sim \pi_\theta(\tau)}[r(\tau)]\\
		&=\mathbb{E}_{\tau\sim \pi_\theta(\tau)}[\nabla_{\theta} \log \pi_\theta(\tau)r(\tau)]\\
		&=\mathbb{E}_{\tau \sim \pi_\theta(\tau)}[(\sum_{t=1}^{T}\nabla_{\theta}\log \pi_\theta(a_t|s_t))(\sum_{t=1}^{T}\gamma^t r(s_t,a_t))]	
	\end{aligned}
\end{equation*}

In deep deterministic policy gradient, four networks are required: online actor, online critic, target actor and target critic. Combining Q-learning and policy gradient, actor is the function $\mu$ and critic is the Q-value function. Agent observe a state and actor would provide an "optimal" action in continuous action space. Then the online critic would evaluate the actor's proposed action and update online actor. What's more, target actor and target critic are used to update online critic.

Formally, the update scheme of DDPG is as below:

For online actor:

\begin{equation*}
	\begin{aligned}
	\nabla_{\theta^{\mu}}J&\approx \mathbb{E}_{s_t\sim \rho^\beta}[\nabla_{\theta^{\mu}}Q(s,a|\theta^Q)|_{s=s_t,a=\mu(s_t|\theta^\mu)}]\\
	&=\mathbb{E}_{s_t\sim \rho^\beta}[\nabla_a Q(s,a|\theta^Q)|_{s=s_t,a=\mu(s_t)}\nabla_{\theta^{\mu}}\mu(s|\theta^\mu)|_{s=s_t}]
	\end{aligned}
\end{equation*}

For online critic, the update rule is similar. The target actor and target critic are updated softly from online actor and online critic. We would leave the details in the presentation of the algorithm:

\begin{algorithm}
	\caption{DDPG}
	\label{algorithm 1}
	\begin{algorithmic}[1]
		\STATE Randomly initialize actor $\mu(s|\theta^{\mu})$ and critic  $Q(s,a|\theta^Q)$
		\STATE Create $Q'$ and $\mu'$ by $\theta^{Q'} \rightarrow \theta^Q$,$\theta^{\mu'}\rightarrow \theta^\mu$
		\STATE Initialize replay buffer \emph{R}
		\FOR{$i=1$ to M} % For 语句，需要和EndFor对应
			\STATE Initialize a UO process $\mathcal{N}$
			\STATE Receive initial observation state $s_1$
			\FOR{$t=1$ to T}
				\STATE Select action $a_t=\mu(s_t|\theta^\mu)+\mathcal{N}_t$
				\STATE Execute action $a_t$ and observe $r_t$ and $s_{t+1}$
				\STATE Save transition ($s_t$,$a_t$,$r_t$,$s_{t+1}$) in \emph{R}
				\STATE Sample a random minibatch of N transitions ($s_i$,$a_i$,$r_i$,$s_{i+1}$) in \emph{R}
				\STATE Set $y_i=r_i+\gamma Q'(s_{i+1},\mu'(s_{i+1}|\theta^{\mu'})|\theta^{Q'})$
				\STATE Update critic by minimizing the loss:$L=\frac{1}{N}\sum_{i}(y_i-Q(s_i,a_i|\theta^Q))^2$
				\STATE Update actor policy by policy gradient:
				\begin{equation*}
					\begin{aligned}
					&\nabla_{\theta^{\mu}}J\\
					&\approx \frac{1}{N}\sum_i\nabla_{\theta^{\mu}}Q(s,a|\theta^Q)|_{s=s_t,a=\mu(s_t|\theta^\mu)}\nabla_{\theta^{\mu}}\mu(s|\theta^\mu)|_{s_t}
					\end{aligned}
				\end{equation*}
				\STATE Update the target networks:
				$$\theta^{Q'}\rightarrow \tau\theta^Q+(1-\tau)\theta^{Q'}$$
				$$\theta^{\mu'}\rightarrow \tau\theta^\mu+(1-\tau)\theta^{\mu'}$$
			\ENDFOR
		\ENDFOR
	\end{algorithmic}
\end{algorithm}

\subsection{Proximal Policy Optimization}

Most algorithms for policy optimization can be classified into three broad categories:(1) policy iteration methods. (2) policy gradient methods and (3) derivative-free optimization methods. Proximal Policy Optimization(PPO) falls into the second category. Since PPO is based on Trust Region Policy Optimization(TRPO) [\ref{TRPO}], we would introduce TRPO first and then PPO.

TRPO finds an lower bound for policy improvement so that policy optimization can deal with surrogate objective function. This could guarantee monotone improvement in policies.

Formally, let $\pi$ denote a stochastic policy $\pi:S\times A \rightarrow [0,1]$, which indicates that the policy would derive a distribution in continuous action space in the given state to represent all the action's fitness. Let
\begin{equation*}
\begin{aligned}
\eta(\pi)=\mathbb{E}_{s_0,a_0,\dots}&[\sum_{t=0}^{\infty}\gamma^tr(s_t)]\\
s_0\sim \rho_0(s_0), a_t\sim \pi(a_t|s_t), &s_{t+1}\sim P(s_{t+1},a_{t+1}|s_{t},a_{t})
\end{aligned}
\end{equation*}

Following standard definitions of the state-action value function $Q_\pi$, the value function $V_\pi$ and the advantage function as below:
$$V_\pi(s_t)=\mathbb{E}_{a_t,s_{t+1},\dots}[\sum_{l=0}^{\infty}\gamma^lr(s_{t+l})]$$

$$Q_\pi(s_t,a_t)=\mathbb{E}_{s_{t+1},a_{t+1},\dots}[\sum_{l=0}^{\infty}\gamma^lr(s_{t+l})]$$

$$A_\pi(s,a)=Q_\pi(s,a)-V_\pi(s)$$

The expected return of another policy $\tilde{\pi}$ over $\pi$ can be expressed in terms of the advantage accumulated over timesteps:

\begin{equation*}
	\begin{aligned}
	\eta(\tilde{\pi})=\eta(\pi)+\mathbb{E}_{s_0,a_0,\dots \sim \tilde{\pi}}[\sum_{t=0}^{\infty}\gamma^t A_\pi(s_t,a_t)]
	\end{aligned}
\end{equation*}

The above equation can be rewritten in terms of states:

\begin{equation*}
	\begin{aligned}
	\eta(\tilde{\pi})&=\eta(\pi)+\sum_{t=0}^{\infty}\sum_{s}P(s_t=s|\tilde{\pi})\sum_{a}\tilde{\pi}(a|s)\gamma^tA_\pi(s,a)\\
	&=\eta(\pi)+\sum_s\sum_{t=0}^{\infty}\gamma^tP(s_t=s|\tilde{\pi})\sum_a\tilde{\pi}(a|s)A_\pi(s,a)\\
	&=\eta(\pi)+\sum_s\rho_{\tilde{\pi}}(s)\sum_a\tilde{\pi}(a|s)A_\pi(s,a)
	\end{aligned}
\end{equation*}

where  $\rho_{\tilde{\pi}}=P(s_0=s)+\gamma P(s_1=s)+\gamma^2 P(s_2=s)+\cdots$ denotes the discounted visitation frequencies of state s given policy $\tilde{\pi}$.

However, the complexity due to the reliance to policy $\tilde{\pi}$ makes the equation difficult to compute. Instead, TRPO proposes the following local approximation.

$$L_\pi(\tilde{\pi})=\eta(\pi)+\sum_s \rho_\pi(s)\sum_a \tilde{\pi}(a|s)A_\pi(s,a)$$

The lower bound of policy improvement, as one of the key results of TRPO, provides theoretical guarantee for monotonic policy improvement:

$$\eta(\pi_{new})\geq L_{\pi_{old}}(\pi_{new})-\frac{4\epsilon\gamma}{(1-\gamma)^2}\alpha^2$$

where
\begin{equation*}
\begin{aligned}
&\epsilon=\max_{s,a}|A_\pi(s,a)|\\	
\alpha&=D_{TV}^{max}(\pi_{old},\pi_{new})\\
&=\max_s D_{TV}(\pi_{old}(\cdot|s)||\pi_{new}(\cdot|s))
\end{aligned}
\end{equation*}

$D_{TV}(p||q)=\frac{1}{2}\sum_i|p_i-q_i|$ is the total variation divergence distance between two discrete probability distributions.

Since $D_{KL}(p||q)\geq D_TV(p||q)^2$, we can derive the following inequation, which is used in the construction of the algorithm:

$$\eta(\tilde{\pi})\geq L_{\pi}(\tilde{\pi})-CD_{KL}^{max}(\pi,\tilde{\pi})$$

where
\begin{equation*}
	\begin{aligned}
	C&=\frac{4\epsilon\gamma}{(1-\gamma)^2}\\
	D_{KL}^{max}(\pi,\tilde{\pi})&=\max_s D_{KL}(\pi(\cdot|s)||\tilde{\pi}(\cdot|s))
	\end{aligned}
\end{equation*}

The proofs of above equations are available in [\ref{TRPO}]

To go further into the detail, let $M_i(\pi)=L_{\pi_i}(\pi)-CD_{KL}^{max}(\pi_i,\pi)$. Two properties would be uncovered without much difficulty as follow:

\begin{equation*}
	\begin{aligned}
	\eta(\pi_i)=M_i(\pi_i)
	\end{aligned}
\end{equation*}

\begin{equation*}
	\begin{aligned}
	\eta(\pi_{i+1})\geq M_i(\pi_{i+1})
	\end{aligned}
\end{equation*}

Therefore, the lower bound of the policy improvement is given out:
$$\eta(\pi_{i+1})-\eta(\pi_i)\geq M_i(\pi_{i+1})-M_i(\pi_i)$$

Thus, by maximizing $M_i$ at each iteration, we guarantee that the true objective $\eta$ is non-decreasing. Consider parameterized policies $\pi_{\theta_i}$, the policy optimization can be turned into:

$$\max_{\pi_{\theta_i}}[L_{\pi_{\theta_{i-1}}}(\pi_{\theta_i})-CD_{KL}^{max}(\pi_{\theta_{i-1}},\pi_{\theta_i})]$$

However, the penalty coefficient C from the theoretical result would provide policy update with too small step sizes. While in the final TRPO algorithm, an alternative optimization problem is proposed after carefully considerations of the structure of the objective function:

\begin{equation*}
	\begin{aligned}
	\max_{\pi_{\theta_i}}& L_{\pi_{\theta_i}}	\\
	s.t. \quad &\overline{D}_{KL}^{\rho\pi_{\theta_{i-1}}}(\pi_{\theta_{i-1}},\pi_{\theta_{i}})\leq\delta\\
	\end{aligned}
\end{equation*}
where $\overline{D}_{KL}^{\rho}(\pi_{\theta_{1}},\pi_{\theta_{2}})=\mathbb{E}_{s\sim \rho}[D_{KL}(\pi_{\theta_{1}}(\cdot|s)||\pi_{\theta_{2}}(\cdot|s))]$

Further approximations are proposed to make the optimization tractable. Recalled that the origin optimization problem can be written as :
$$\max_{\pi_{\theta}} \sum_s \rho \pi_{\theta_{i-1}}(s)\sum_a \pi_{\theta_{i}}(a|s)A_{\theta_{i-1}}(s,a)$$

After some approximations including importance sampling, the final optimization comes into:

$$\max_{\pi_{\theta_i}}\mathbb{E}_{s\sim \rho \pi_{\theta_{i-1}},a\sim q}[\frac{\pi_{\theta_{i}}(a|s)}{q(a|s)}A_{\pi_{\theta_{i-1}}}(s,a)]$$
$$s.t. \quad \mathbb{E}_{s \sim \rho \pi_{\theta_{i-1}}}[D_{KL}(\pi_{\theta_{i-1}}(\cdot|s)||\pi_{\theta_{i}}(\cdot|s))]\leq \delta$$

So here comes the PPO [\ref{PPO}]: it proposed new surrogate objective to simplify TRPO. One of them is clipped surrogate objective which we choose in our experiments. Let us denote $r(\theta)=\frac{\pi_\theta(a|s)}{\pi_{\pi_{old}}(a|s)}$. The clipped surrogate objective can be written as:

$$L^{CLIP}(\theta)=\mathbb{E}[\min(r(\theta)A,clip(r(\theta),1-\epsilon,1+\epsilon)A)]$$

This net surrogate objective function can constrain the update step in a much simpler manner and experiments show it does outperform the original objective function in terms of sample complexity.

\begin{algorithm}[H]
	\caption{PPO}
	\begin{algorithmic}[1]
		\STATE Initialize actor $\mu:S \rightarrow \mathbb{R}^{m+1}$ and \\ $\sigma:S \rightarrow diag(\sigma_1,\sigma_2,\cdots,\sigma_{m+1})$
		\FOR{$i=1$ to M} % For 语句，需要和EndFor对应
			\STATE Run policy $\pi_\theta\sim N(\mu(s),\sigma(s))$ for T timesteps and collect $(s_t,a_t,r_t)$
			\STATE Estimate advantages $\hat{A_t}=\sum_{t'>t}\gamma^{t'-t}r_{t'}-V(s_t)$
			\STATE Update old policy $\pi_{old} \leftarrow \pi_\theta$
			\FOR{$j=1$ to N}
				\STATE Update actor policy by policy gradient:
				$$\sum_i \nabla_{\theta} L_i^{CLIP}(\theta)$$
				\STATE Update critic by:
				$$\nabla L(\phi)=-\sum_{t=1}^{T} \nabla \hat{A_t}^2$$
				
			\ENDFOR
		\ENDFOR
	\end{algorithmic}
\end{algorithm}

\section{Adversarial Learning}

Although deep reinforcement learning is potential in portfolio management for its competence in capturing nonlinear features and low prior assumption and similarity with human investing, three main characteristics are worth to pay attention:

\begin{itemize}
    \item Financial market is highly volatile and non-stationary, which is totally different with game or robot control
    \item Conventional reinforcement learning is designed for infinite-horizon MDP while portfolio management seeks to maximize absolute portfolio value or other objective in finite time
    \item In game playing or robot control there is no need for splitting training set and testing set while in financial market a satisfying performance in back test is essential in evaluating strategies
    \item Stock market has explicit expression for portfolio value which does not exist in game playing and robot control. Therefore, approximating the value function is useless and can even deteriorate agent's performance due to the difficulty and error in approximation
\end{itemize}

Therefore, some modifications are needed in order to apply this method in portfolio management. Adapting average return instead of discount return can mitigate the contradiction between infinite and finite horizon. In our experiments, we found that DDPG and PPO both have unsatisfying performance in training process, indicating that they cannot figuring out the optimal policy even in training set.

As for the higher robust and risk-sensitive requirement for deep reinforcement learning in portfolio management, we propose so called adversarial training. In fact, risk-sensitive MDP and robust MDP are both two preferable method especially in portfolio management. LA Prashanth et al.  devise actor-critic algorithms for estimating the gradient and updating the policy parameters in the ascent direction while establishing the convergence of our algorithms to locally risk-sensitive optimal policies [\ref{AC-risk}]. Motivated by A Pattanaik et al., who train two reinforcement learning agents and adversarial playing for enhancing main player robustness [\ref{adversarial}], and $L_{\infty}$ control, we propose adversarial training, which is to adding random noise in market prices. In our experiments, we add $N(0,0.002)$ noise in the data. However, based on Conditional Value at risk (CVaR), Non-zero expectation distribution can also be adapted to make agents more conservative.

$$CVaR=\frac{1}{1-c}\int_{-1}^{VaR}xp(x)dx$$

Therefore, we give out our revised Policy Gradient in our following experiments as below:
\begin{algorithm}
	\caption{Adversarial PG}
	\label{algorithm 3}
	\begin{algorithmic}[1]
		\STATE Randomly initialize actor $\mu(s|\theta^{\mu})$
		\STATE Initialize replay buffer \emph{R}
		\FOR{$i=1$ to M} % For 语句，需要和EndFor对应
			\STATE Receive initial observation state $s_1$
            \STATE Add noise into the price data
			\FOR{$t=1$ to T}
				\STATE Select action $\omega_t=\mu(s_t|\theta^\mu)$
				\STATE Execute action $\omega_t$ and observe $r_t$, $s_{t+1}$ and $\omega_t'$
				\STATE Save transition ($s_t$,$\omega_t$,$\omega_t'$) in \emph{R}
			\ENDFOR
            \STATE Update actor policy by policy gradient:
            				\begin{equation*}
            					\begin{aligned}
            					&\nabla_{\theta^{\mu}}J
            					= \nabla_{\theta^{\mu}} \frac{1}{N}\sum_{t=1}^T(\log(\mathbf{\omega_{t}} \cdot \mathbf{y_{t}}-\mu \sum_{i=1}^{m}|\omega_{i,t}-\omega'_{i,t-1}|)
            					\end{aligned}
            				\end{equation*}
		\ENDFOR
	\end{algorithmic}
\end{algorithm}

\section{Experiments}

\subsection{Data preparation}
Our experiments are conducted on China Stock data from investing\footnote{https://lpi.invest.com/invest.com\&bittrex}, wind\footnote{http://www.wind.com.cn/}. A fixed number (which is 5 in our experiments) of assets are randomly chosen from the assets pool. To ensure enough data is provided for learning, after a portfolio is formed, we check the intersection of their available trading history and only if it is longer than our pre-set threshold (which is 1200 days) can we run our agent on it.

In order to derive a general agent which is robust with different stocks, we normalize the price data. To be specific, we divide the opening price, closing price, high price and low price by the close price at the last day of the period. For missing data which occurs during weekends and holidays, in order to maintain the time series consistency, we fill the empty price data with the close price on the previous day and we also set volume 0 to indicate the market is closed at that day.

\subsection{network structure}
Motivated by Jiang et al., we use so called Identical Independent Evaluators(IIE). IIE means that the networks flow independently for the m+1 assets while network parameters are shared among these streams. The network evaluates one stock at a time and output a scaler to represent its preference to invest in this asset. Then m+1 scalers are normalized by softmax function and compressed into a weight vector as the next period's action. IIE has some crucial advantages over an integrated network, including scalability in portfolio size, data-usage efficiency and plasticity to asset collection. The explanation can be reviewed in [\ref{DRLPM}] and we are not going to illustrate them here.

We find that in other works about deep learning in portfolio management, CNN outperforms RNN and LSTM in most cases. However, different from Jiang et al., we alternate CNN with Deep Residual Network. The depth of the neural network plays an important role in its performance. However, conventional CNN network is stopped from going deeper because of gradient vanishment and gradient explosion when the depth of the networks increases. Deep residual network solves this problem by adding a shortcut for layers to jump to the deeper layers directly, which could prevent the network from deteriorating as the depth adds. Deep Residual Network has gained remarkable performance in image recognition and greatly contributes to the development of deep learning.[\ref{res}] When it comes to our structure of PG, we adapt similar settings with Jiang's and we would not go specific about them here.

\begin{figure}[ht]
	\centering
	\includegraphics[scale=0.3]{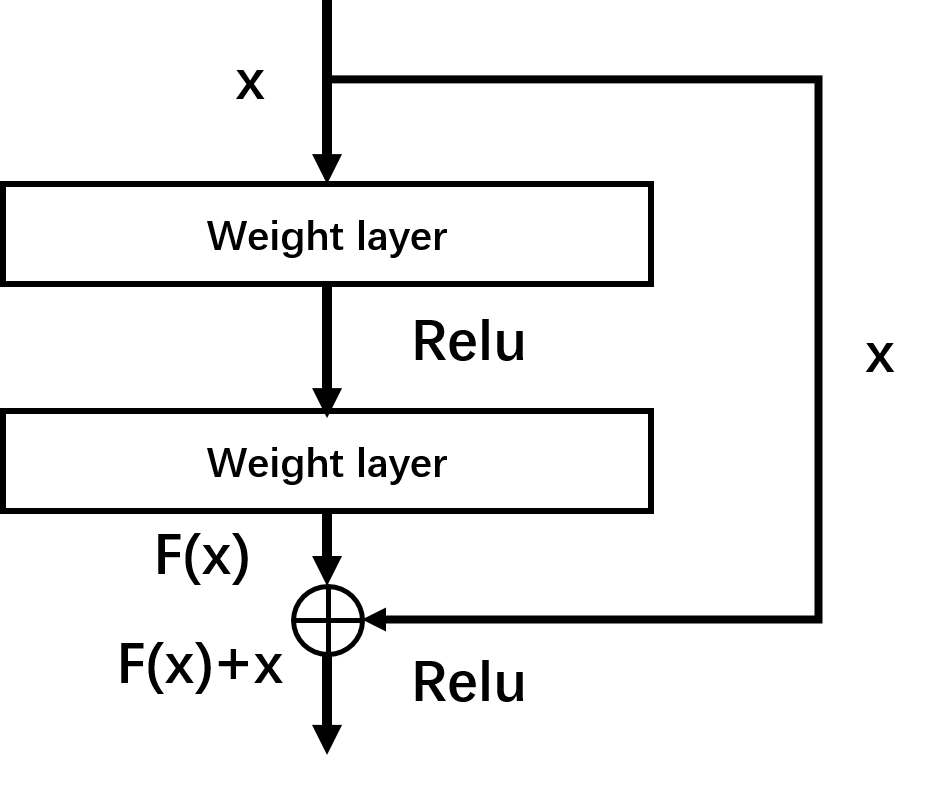}
	\caption{Residual Block} \label{resnet}
\end{figure}

\begin{figure}[ht]
	\centering
	\includegraphics[scale=0.26]{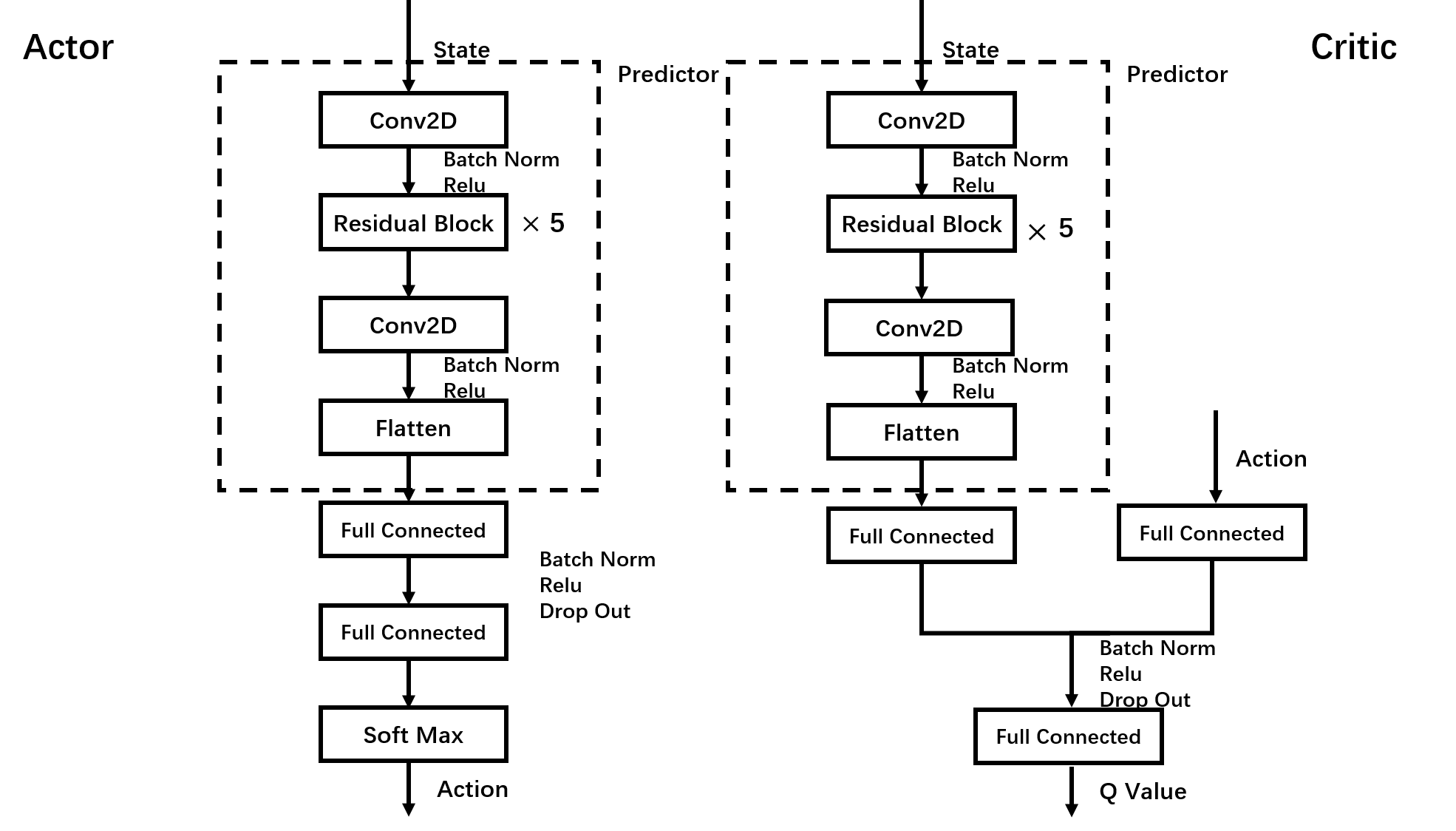}
	\caption{DDPG Network Structure in our experiments} \label{DDPG structure}
\end{figure}

\begin{figure}[ht]
	\centering
	\includegraphics[scale=0.26]{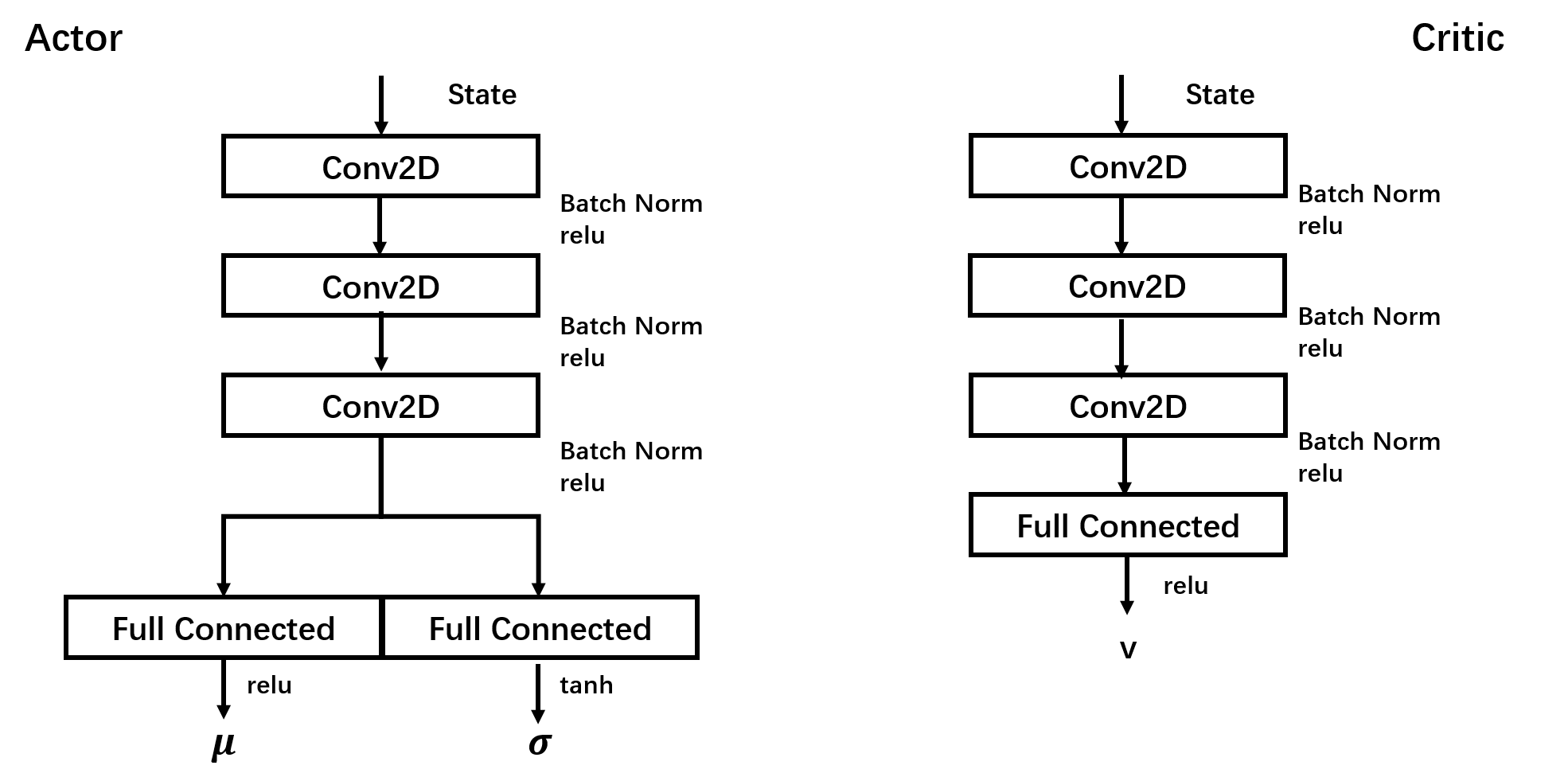}
	\caption{PPO Network Structure in our experiments} \label{PPO structure}
\end{figure}

\begin{table}[H]
	\centering
	\begin{tabular}{*{5}{c}}
		\bottomrule
		\multirow{2}*{Algorithm} & \multicolumn{2}{c}{DDPG} & \multicolumn{2}{c}{PPO}\\
		\cmidrule(lr){2-3}\cmidrule(lr){4-5}
		& Actor & Critic & Actor & Critic\\
		\midrule
		Optimizer & Adam & Adam & GradientDescent & GradientDescent\\
		Learning Rate & $10^{-3}$ & $10^{-1}$ & $10^{-3}$ & $10^{-3}$\\
		$\tau$ & $10^{-2}$  & $10^{-2}$  & $10^{-2}$  & $10^{-2}$ \\
		\bottomrule
	\end{tabular}
	\vspace{0.5cm}
	\caption{Hyper parameters in our experiments}
\end{table}

\subsection{result}

\subsubsection{learning rate}

Learning rate plays an essential role in neural network training. However, it is also very subtle. A high learning rate will make training loss decrease fast at the beginning but drop into a local minimum occasionally, or even vibrate around the optimal solution but could not reach it. A low learning rate will make the training loss decrease very slowly even after a large number of epochs. Only a proper learning rate can help network achieve a satisfactory result.

Therefore, we implement DDPG and test it using different learning rates. The results show that learning rates have significant effect on critic loss even actor's learning rate does not directly control the critic's training. We find that when the actor learns new patterns, critic loss would jump. This indicates that the critic has not sufficient generalization ability towards new states. Only when the actor becomes stable can the critic loss decreases.

\begin{figure}[h]
	\centering
	\includegraphics[scale=0.41]{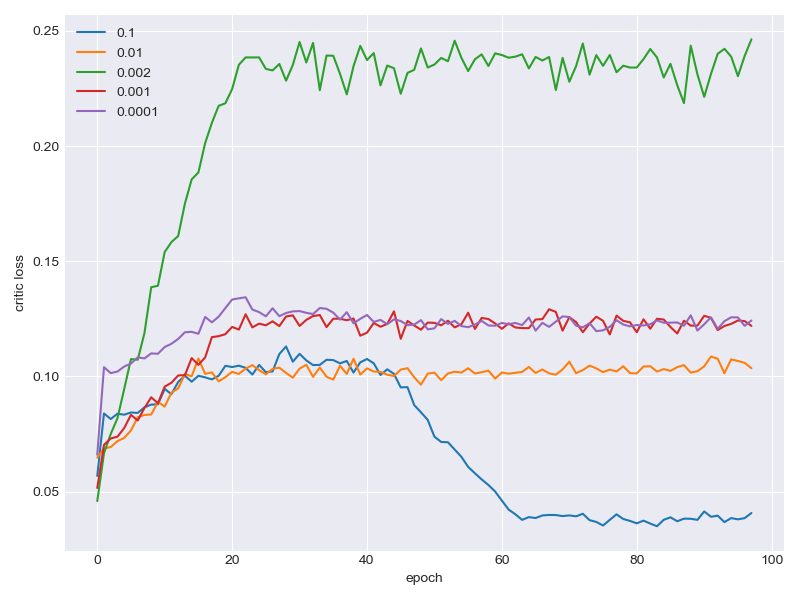}
	\caption{Critic loss under different actor learning rates} \label{critic_loss(AL)}
\end{figure}

\begin{figure}[h]
	\centering
	\includegraphics[scale=0.41]{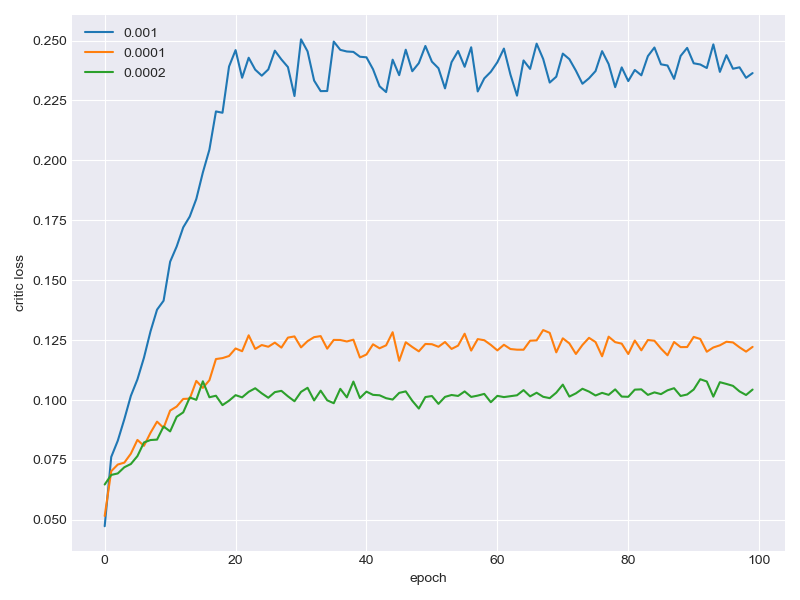}
	\caption{Critic loss under different critic learning rates} \label{critic_loss(CL)}
\end{figure}

\subsubsection{Risk}

Due to the limitation of training data, our reinforcement learning agent may underestimate the risk when training in bull market, which may occur disastrous deterioration in its performance in real trading environment. Different approaches in finance can help evaluate the current portfolio risk to alleviate the effect of biased training data. Inspired by Almahdi et al. in which objective function is risk-adjusted and Jacobsen et al. which shows the volatility would cluster in a period, we modify our objective function as follow:

$$R=\sum_{t=1}^{T}\gamma^t(r(s_t,a_t)-\beta \sigma^2_t)$$
where $\sigma_t^2=\frac{1}{L}\sum_{t'=t-L+1}^{t}\sum_{i=1}^{m+1}(y_{i,t'}-\overline{y_{i,t'}})^2\cdot w_{i,t}$ and $\overline{y_{i,t'}}=\frac{1}{L}\sum_{t'=t-L+1}^{t}y_{i,t'}$ measure the volatility of the returns of asset i in the last L day. The objective function is constrained by reducing the profit from investing in highly volatile assets which would make our portfolio exposed in exceeded danger.

\begin{figure}[h]
	\label{risk}
	\centering
	\includegraphics[scale=0.41]{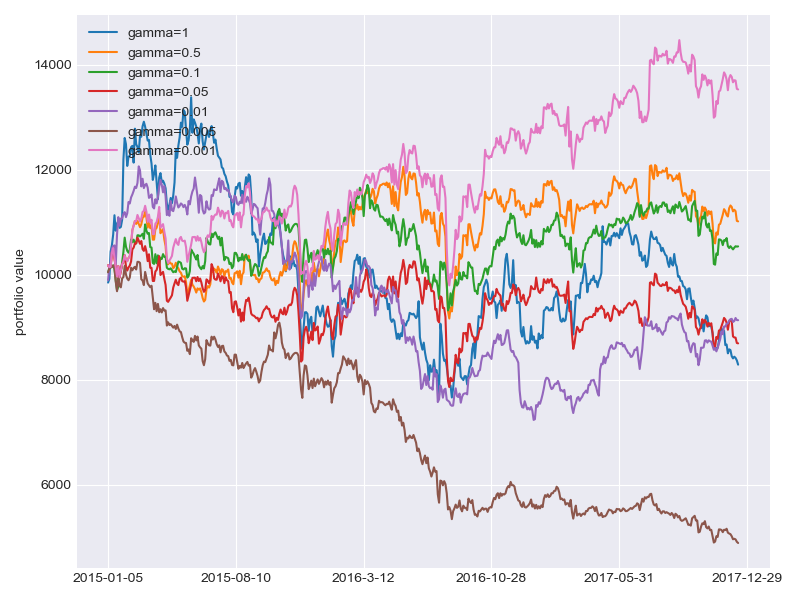}
	\caption{Comparison of portfolio value with different risk penalties($\beta$)}
\end{figure}

Unfortunately, the result seems not support our modifications. We also train our agent in objective function taking form of Sharpe ratio but it also fails. In fact, reward engineering is one of the core topics in designing reinforcement learning algorithms [\ref{reward eng}]. It seems that our modification makes the objective function too complex.

\subsubsection{Features combination}
As far as we know, few works discuss the combinations of features in reinforcement learning. Different from end to end game playing or robot control whose input is pixels, in portfolio management, abundant features can be taken into considerations. Common features include the closing price, the open price, the high price, the low price and volume. What's more, financial indexes for long term analysis such as Price-to-Earning Ratio (PE), Price to book ratio (PB) can also provide insights into market movements.

However, adding irrelevant features would add noise and deteriorate the training. The trade off in it is the topic of feature selection. Therefore, we conduct experiments under different combinations of features, which are 1. only with closing prices, 2. with closing and high, 3. with closing and open, 4. with closing and low prices. The results show that feature combinations matter in the training process. Select closing and high price could help agent gain the best performance in our experiments.

\begin{figure}[ht]
	\label{features_critic_loss}
	\centering
	\includegraphics[scale=0.41]{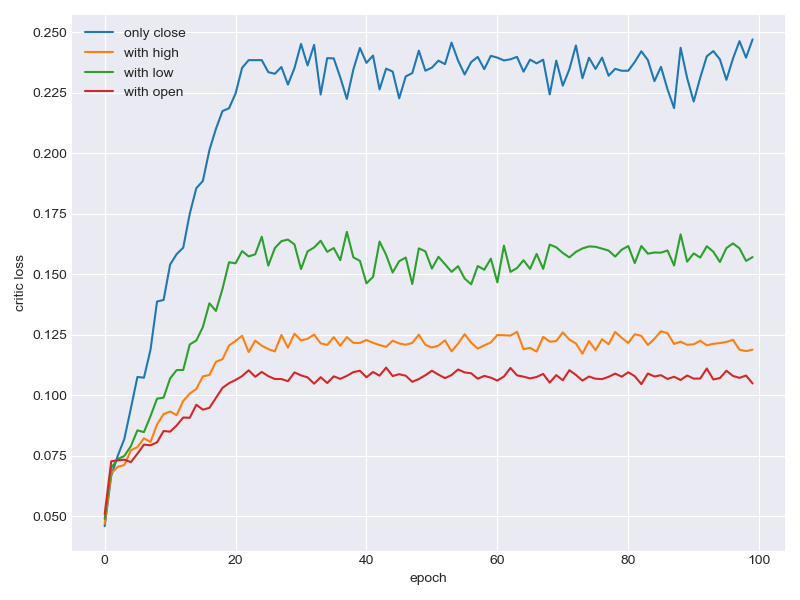}
	\caption{Comparison of critic loss with different features combinations}
\end{figure}

\begin{figure}[ht]
	\label{features_reward}
	\centering
	\includegraphics[scale=0.41]{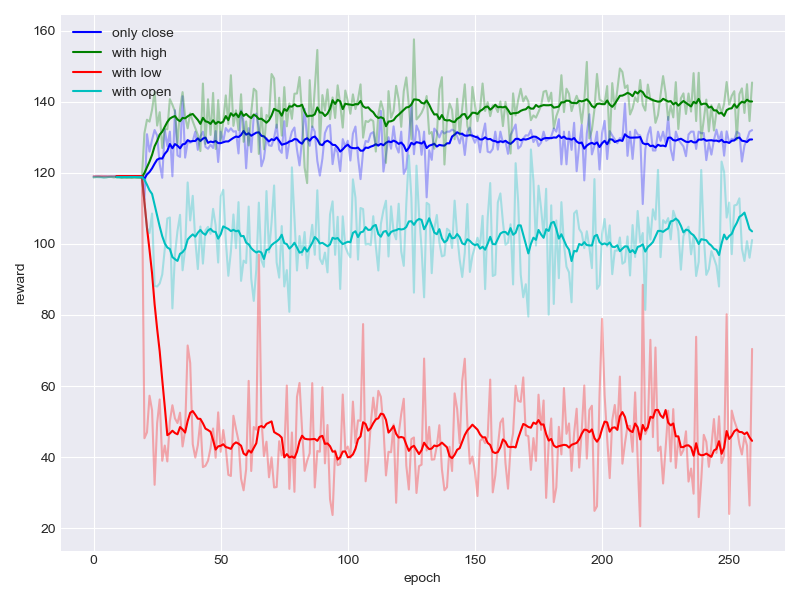}
	\caption{Comparison of reward with different features combinations}
\end{figure}

\subsubsection{Training and Testing}

After experiments mentioned above, we derive a satisfying set of hyper parameters and features combinations. Under such setting, we conduct training for 1000 epochs on both China stock market and USA stock market. The result shows that training could increase accumulative portfolio value (APV) while reducing the volatility of the returns in training data.

\begin{figure}[ht]
	\centering
	\includegraphics[scale=0.41]{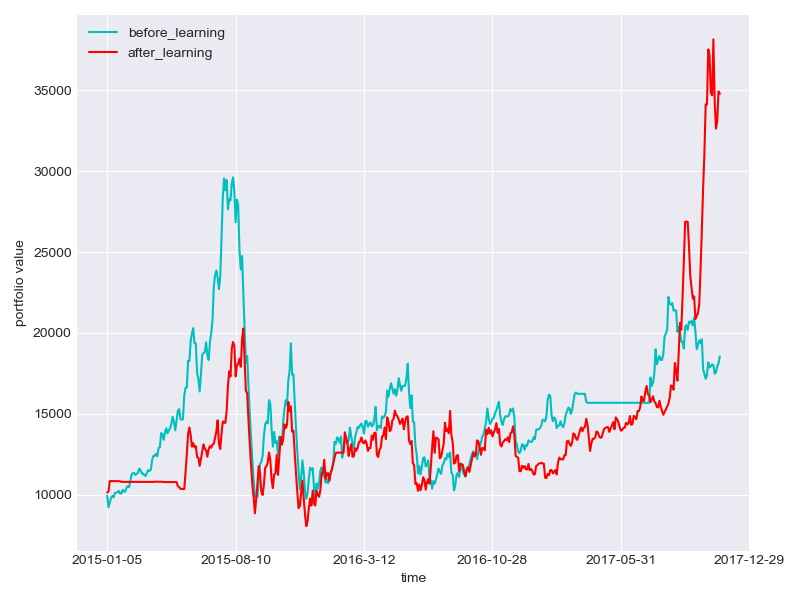}
	\caption{Comparison of portfolio value before and after learning in training data of China stock market by DDPG} \label{comparison}
\end{figure}

\begin{figure}[ht]
	\centering
	\includegraphics[scale=0.41]{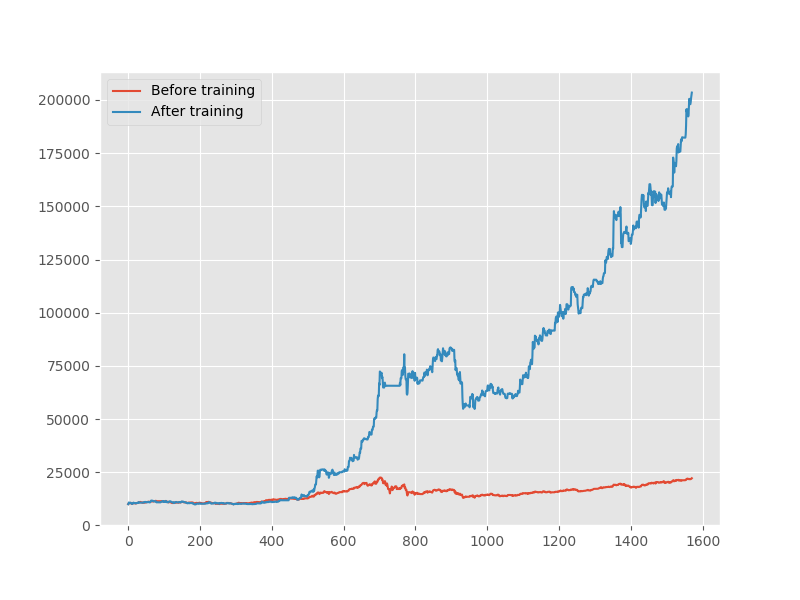}
	\caption{Comparison of portfolio value before and after learning in training data of China stock market by PG} \label{comparison2}
\end{figure}

\subsubsection{Noise}

\begin{figure}[ht]
	\centering
	\includegraphics[scale=0.41]{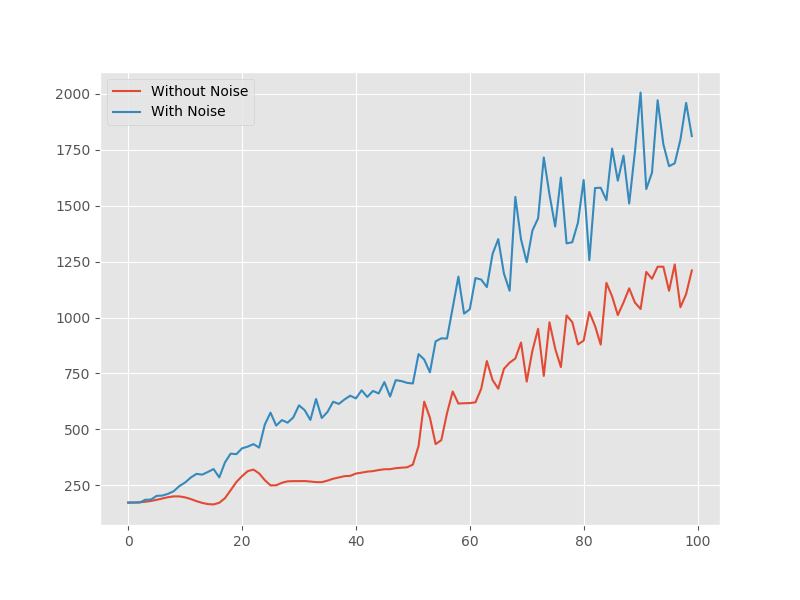}
	\caption{Comparison of portfolio value in training process with and without adversarial learning} \label{comparison3}
\end{figure}

We present the training process using adversarial learning and without using it: [\ref{comparison3}]

We conduct 50 groups of experiments and we show 25 of them below.

\begin{table}[h]
	\centering
	\begin{tabular}{lllllll}
		\toprule
		&  ADR(\%) & Sharpe & MMD & ADR(\%) & Sharpe & MMD\\
		\midrule
		1	& 0.416	& 1.171	& 0.416	& 0.226	& 0.678	& 0.22\\
		2	& 0.242	& 0.647	& 0.417	& 0.31	& 0.885	& 0\\
		3	& 0.242	& 0.724	& 0.224	& 0.249	& 0.753	& 0.13\\
		4	& 0.298	& 0.859	& 0.349	& 0.304	& 0.921	& 0.119\\
		5	& 0.262	& 0.765	& 0.45	& 0.254	& 0.802	& 0\\
		6	& 0.413	& 1.142	& 0.305	& 0.323	& 0.903	& 0\\
		7	& 0.213	& 0.668	& 0.449	& 0.202	& 0.667	& 0.231\\
		8	& 0.187	& 0.554	& 0.347	& 0.276	& 0.836	& 0\\
		9	& 0.471	& 1.107	& 0.649	& 0.308	& 0.873	& 0.277\\
		10	& 0.32	& 0.795	& 0.546	& 0.297	& 0.812	& 0.279\\
		11	& 0.312	& 0.837	& 0.195	& 0.338	& 0.924	& 0\\
		12	& 0.17	& 0.741	& 0.242	& 0.202	& 0.715	& 0.19\\
		13	& 0.313	& 0.825	& 0.341	& 0.26	& 0.691	& 0.34\\
		14	& 0.345	& 0.931	& 0.263	& 0.307	& 0.892	& 0.096\\
		15	& 0.573	& 1.499	& 0.609	& 0.354	& 0.993	& 0.272\\
		16	& 0.493	& 1.337	& 0.42	& 0.328	& 0.91	& 0\\
		17	& 0.348	& 0.911	& 0.307	& 0.364	& 1.002	& 0.23\\
		18	& 0.198	& 0.601	& 0.251	& 0.244	& 0.756	& 0.086\\
		19	& 0.306	& 0.813	& 0.46	& 0.295	& 0.863	& 0\\
		20	& 0.377	& 1.099	& 0.419	& 0.313	& 0.949	& 0.165\\
		21	& 0.325	& 0.876	& 0.23	& 0.308	& 0.828	& 0.237\\
		22	& 0.301	& 0.918	& 0.123	& 0.269	& 0.824	& 0\\
		23	& 0.373	& 1.176	& 0.514	& 0.245	& 0.819	& 0.138\\
		24	& 0.487	& 1.257	& 0.461	& 0.33	& 0.914	& 0\\
		25	& 0.426	& 1.14	& 0.386	& 0.415	& 1.127	& 0.408\\
		\bottomrule
	\end{tabular}
	\vspace{0.5cm}
	\caption{Performances of Adversarial and not-adversarial Learning}
\end{table}

After adding noise in the prices data of stocks, remarkable progress can be seen in the training process with successfully avoiding sticking on saddle points and significantly much higher APV after 100 epochs training.
We also conduct statistic test on the backtest result. Our first null  hypothesis is:
$$H_0:\quad ARR_1<ARR_2$$

The p value of the t test is 0.0076 and we can have at least 99\% confidence to believe this adversarial training process can improve the average daily return.

What's more, we want to investigate whether this modification can improve sharpe ratio in the backtest. Our second null hypothesis is :
$$H_0:\quad SharpeRatio_1<SharpeRatio_2$$

The p value of the t test is 0.0338 and we can have at least 95\% confidence to believe this modifications indeed promote the sharpe ratio.

Finnaly, we want to investigate whether this modification would make our agent more volatile. Our third null hypothesis is :
$$H_0:\quad MaxMarkDown_1<MaxMarkDown_2$$

The p value of the t test is 2.73e-8 and we can have at least 99.9\% confidence to believe this modifications indeed make our agent more volatile.

Then we back test our agent on  China data. The unsatisfying performance of PPO algorithm uncovers the considerable gap between game playing or robot control and portfolio management. Random policy seems unsuitable in such an unstationary, low signal noise ratio financial market although its theoretical properties are appealing, including monotone improvement of policy and higher sample efficiency. All the detail experiments results, including the portfolio stocks codes, training time period, testing time period and average daily return, sharpe ratio and max drawdown of PG agent, URCP agent, Follow-the-winner and Follow-the-loser agent can be viewed in \footnote{https://github.com/qq303067814/Reinforcement-learning-in-portfolio-management-/tree/master/Experiments\%20Result}.

\begin{figure}[ht]
	\centering
	\includegraphics[scale=0.41]{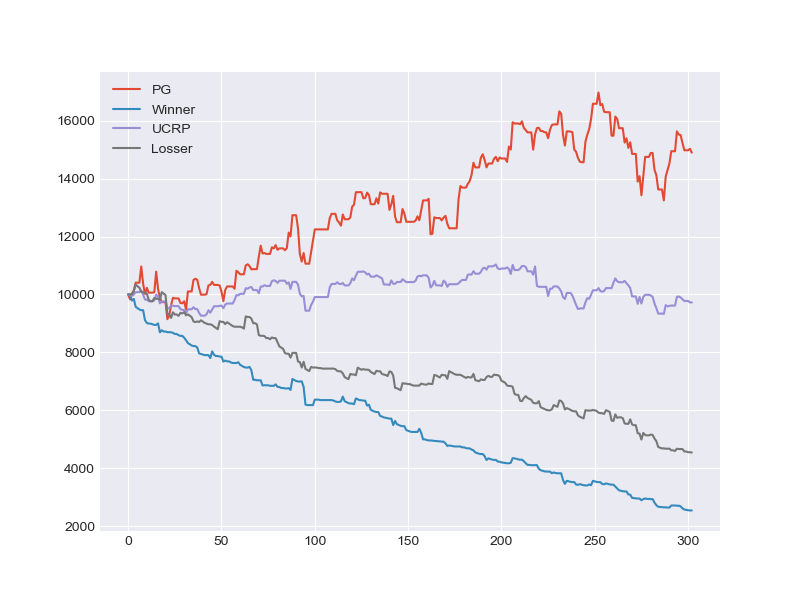}
	\caption{Backtest on China Stock Market} \label{Final}
\end{figure}

$$H_0:\quad ARR_1<ARR_2$$
The p value of the t test is 0.039 and we can have at least 95\% confidence to believe PG agent can outperform URCP on average daily return.

$$H_0:\quad SharpeRatio_1<SharpeRatio_2$$
The p value of the t test is 0.013 and we can have at least 95\% confidence to believe PG agent can outperform URCP on Sharpe Ratio.

$$H_0:\quad MaxMarkDown_1<MaxMarkDown_2$$
The p value of the t test is 1e-11 and we can have at least 99.9\% confidence to believe PG agent's max markdown is higher than URCP.

\section{Future work}
Thanks to the characteristics of portfolio management, there is still many interesting topics in combination with deep reinforcement learning. For future research, we will try to use other indicators to measure the risk of our asset allocation, and work on the combination with conventional models in finance to make advantages of previous finance research. To be specific, we believe model-based reinforcement as a good candidate in portfolio management instead of model-free [\ref{Model based DQN}] [\ref{Model based-Model free}].In model-based reinforcement learning, a model of the
dynamics is used to make predictions, which is used for
action selection. Let $f_{\theta}(s_t; a_t)$ denote a learned discrete-time dynamics function, parameterized by $\theta$, that takes the current
state $s_t$ and action $a_t$ and outputs an estimate of the next
state at time $t + \Delta t$. We can then choose actions by solving the following optimization problem:
$$(a_t,\dots,a_{t+H-1})=\arg \max_{a_t,\dots,a_{t+H-1}} \sum_{t'=t}^{t+H-1}\gamma^{t'-t}r(s_{t'},a_{t'})$$

 What's more, due to the fact that neural network is sensitive to the quality of data, traditional financial data noise reduction approaches can be utilized, such as wavelet analysis [\ref{wavelet}] and the Kalman Filter [\ref{Kalman}]. A different approach for data preprocessing is to combine HMM with reinforcement learning, which is to extract the states beneath the fluctuated prices and learning directly from them [\ref{Bottleneck}].

 Modification of the object function can also be taken into considerations. One direction is to adapt risk-adjust return. Another direction we come up with experiments in designing RL agent in game playing. In game playing, the reward function is simple, for example, in flappy bird, the agent would receive reward 1 when passing a pillar or receive reward -1 when drop to the ground. Complex objective function would hinder agent from achieving desirable performance. We have conduct a naive version of accumulative portfolio value as object function, which is to take win rate instead of absolute return but we cannot receive satisfying improvement.

\section{Conclusion}
This paper applies deep reinforcement learning algorithms with continuous action space to asset allocation. We compare the performances of DDPG ,PPO and PG algorithms in hyper parameters and so on. Compared with previous works of portfolio management using reinforcement learning, we test our agents with risk-adjusted accumulative portfolio value as objective function and different features combinations as input. The experiments show that the strategy obtained by PG algorithm can outperform UCRP in assets allocation. It's found that deep reinforcement learning can somehow capture patterns of market movements even though it is allowed to observe limited data and features and self-improve its performance.

However, reinforcement learning does not gain such remarkable performance in portfolio management so far as those in game playing or robot control. We come up with a few ideas.

First, the second-order differentiability for the parameters in the neural network of the output strategy and the expectation in Q-value is necessary for convergence of the algorithm. Due to the algorithm, we could only search for optimal policy in the second-order differentiable strategy function set, instead of in the policy function set, which might also lead to the failure of finding globally optimal strategy.

Second, the algorithm requires stationary transition. However, due to market irregularities and government intervention, the state transitions in stock market might be time varying.

In our experiments, deep reinforcement learning is highly sensitive so that its performance is unstable. What's more, the degeneration of our reinforcement learning agent, which often tends to buy only one asset at a time, indicates more modifications are needed for designing promising algorithms.

\section*{Acknowledgment}
We would like to say thanks to Mingwen Liu from ShingingMidas Private Fund, Zheng Xie and Xingyu Fu from Sun Yat-sen University for their generous guidance. Without their support, we could not overcome so many challenges during this project.


\begin{thebibliography}{1}
\bibitem{IEEEhowto:kopka}
Li, Bin, and Steven CH Hoi. "Online portfolio selection: A survey." ACM Computing Surveys (CSUR) 46.3 (2014): 35. \label{Online portfolio selection}

\bibitem{IEEEhowto:kopka}
Du, Xin, Jinjian Zhai, and Koupin Lv. "Algorithm trading using q-learning and recurrent reinforcement learning." positions 1 (2009): 1.\label{Q&RRL}

\bibitem{IEEEhowto:kopka}
Almahdi, Saud, and Steve Y. Yang. "An adaptive portfolio trading system: A risk-return portfolio optimization using recurrent reinforcement learning with expected maximum drawdown." Expert Systems with Applications 87 (2017): 267-279.\label{EMDD}

\bibitem{IEEEhowto:kopka}
Gao, Xiu, and Laiwan Chan. "An algorithm for trading and portfolio management using Q-learning and sharpe ratio maximization." Proceedings of the international conference on neural information processing. 2000.
\label{Sharpe Ratio Maximization}

\bibitem{IEEEhowto:kopka}
Li, Yuxi. "Deep reinforcement learning: An overview." arXiv preprint arXiv:1701.07274 (2017).
\label{DRL}

\bibitem{IEEEhowto:kopka}
Sutton, Richard S., et al. "Policy gradient methods for reinforcement learning with function approximation." Advances in neural information processing systems. 2000.
\label{PG with appro}

\bibitem{IEEEhowto:kopka}
Mnih, Volodymyr, et al. "Human-level control through deep reinforcement learning." Nature 518.7540 (2015): 529.\label{DQN}


\bibitem{IEEEhowto:kopka}
Lillicrap, Timothy P., et al. "Continuous control with deep reinforcement learning." arXiv preprint arXiv:1509.02971 (2015).\label{DDPG}


\bibitem{IEEEhowto:kopka}
Schulman, John, et al. "Proximal policy optimization algorithms." arXiv preprint arXiv:1707.06347 (2017). \label{PPO}


\bibitem{IEEEhowto:kopka}
Jiang, Zhengyao, Dixing Xu, and Jinjun Liang. "A deep reinforcement learning framework for the financial portfolio management problem." arXiv preprint arXiv:1706.10059 (2017). \label{DRLPM}

\bibitem{IEEEhowto:kopka}
He K, Zhang X, Ren S, et al. Deep residual learning for image recognition[C]//Proceedings of the IEEE conference on computer vision and pattern recognition. 2016: 770-778. \label{res}

\bibitem{IEEEhowto:kopka}
Guo, Yifeng, et al. "Robust Log-Optimal Strategy with Reinforcement Learning." arXiv preprint arXiv:1805.00205 (2018). \label{RLORL}

\bibitem{IEEEhowto:kopka}
Tang, Lili. "An actor-critic-based portfolio investment method inspired by benefit-risk optimization." Journal of Algorithms and Computational Technology (2018): 1748301818779059.
\label{AC}

\bibitem{IEEEhowto:kopka}
Lu, David W. "Agent Inspired Trading Using Recurrent Reinforcement Learning and LSTM Neural Networks." arXiv preprint arXiv:1707.07338 (2017).
\label{LSTM}

\bibitem{IEEEhowto:kopka}
Yang, Steve Y., Yangyang Yu, and Saud Almahdi. "An investor sentiment reward-based trading system using Gaussian inverse reinforcement learning algorithm." Expert Systems with Applications 114 (2018): 388-401.
\label{inverse}

\bibitem{IEEEhowto:kopka}
Buehler, Hans, et al. "Deep hedging." (2018).
\label{DeepHedging}

\bibitem{IEEEhowto:kopka}
Prashanth, L. A., and Mohammad Ghavamzadeh. "Actor-critic algorithms for risk-sensitive MDPs." Advances in neural information processing systems. 2013.
\label{AC-risk}

\bibitem{IEEEhowto:kopka}
Silver, David, et al. "Deterministic policy gradient algorithms." ICML. 2014.
\label{DPG}

\bibitem{IEEEhowto:kopka}
John Schulman, Sergey Levine, Philipp Moritz, Michael Jordan, Pieter Abbel: Trust Region Policy Optimization \label{TRPO}

\bibitem{IEEEhowto:kopka}
Pattanaik, Anay, et al. "Robust deep reinforcement learning with adversarial attacks." Proceedings of the 17th International Conference on Autonomous Agents and MultiAgent Systems. International Foundation for Autonomous Agents and Multiagent Systems, 2018.
\label{adversarial}

\bibitem{IEEEhowto:kopka}
Jacobsen, Ben, and Dennis Dannenburg. "Volatility clustering in monthly stock returns." Journal of Empirical Finance 10.4 (2003): 479-503.\label{VC}

\bibitem{IEEEhowto:kopka}
Dewey, Daniel. "Reinforcement learning and the reward engineering principle." 2014 AAAI Spring Symposium Series. 2014.\label{reward eng}

\bibitem{IEEEhowto:kopka}
Gu, Shixiang, et al. "Continuous deep q-learning with model-based acceleration." International Conference on Machine Learning. 2016.\label{Model based DQN}

\bibitem{IEEEhowto:kopka}
Nagabandi, Anusha, et al. "Neural network dynamics for model-based deep reinforcement learning with model-free fine-tuning." arXiv preprint arXiv:1708.02596 (2017).\label{Model based-Model free}

\bibitem{IEEEhowto:kopka}
Rua, António, and Luís C. Nunes. "International comovement of stock market returns: A wavelet analysis." Journal of Empirical Finance 16.4 (2009): 632-639.\label{wavelet}

\bibitem{IEEEhowto:kopka}
Faragher, Ramsey. "Understanding the basis of the Kalman filter via a simple and intuitive derivation." IEEE Signal processing magazine 29.5 (2012): 128-132.\label{Kalman}

\bibitem{IEEEhowto:kopka}
Serban, Iulian Vlad, et al. "The Bottleneck Simulator: A Model-based Deep Reinforcement Learning Approach." arXiv preprint arXiv:1807.04723 (2018).\label{Bottleneck}

\bibitem{IEEEhowto:kopka}
Rao, Anil V. "A survey of numerical methods for optimal control." Advances in the Astronautical Sciences 135.1 (2009): 497-528.\label{optimal control}

\end{thebibliography}
\end{document}